\documentclass{vgtc}                          

\ifpdf
  \pdfoutput=1\relax                   
  \usepackage{graphicx}                
  \DeclareGraphicsExtensions{.pdf,.png,.jpg,.jpeg} 
\else
  \usepackage{graphicx}                
  \DeclareGraphicsExtensions{.eps}     
\fi%

\graphicspath{{figures/}{pictures/}{images/}{./}} 
  \usepackage{graphicx}                
\usepackage{microtype}                 
\PassOptionsToPackage{warn}{textcomp}  
\usepackage{textcomp}                  
\usepackage{mathptmx}                  
\usepackage{times}                     
\usepackage{cite}                      
\usepackage{tabu}                      
\usepackage{booktabs}                  

\usepackage{subcaption}
\usepackage[usenames,dvipsnames,svgnames,table]{xcolor}

\onlineid{0}

\vgtccategory{Research}

\title{Evaluating Semantic Interaction on Word Embeddings via Simulation}

\author{Yali Bian\thanks{e-mail: yali@vt.edu}\\ %
        \scriptsize Virginia Tech %
\and Michelle Dowling\thanks{e-mail:dowlingm@vt.edu}\\%
       \scriptsize Virginia Tech %
\and Chris North\thanks{e-mail:north@vt.edu}\\ %
     \scriptsize Virginia Tech %
     }

\abstract{
Semantic interaction (\textit{SI}) attempts to learn the user's cognitive intents as they directly manipulate data projections during sensemaking activity.  
For text analysis, prior implementations of SI have used common data features, such as bag-of-words representations, for machine learning from user interactions.  Instead, we hypothesize that features derived from deep learning word embeddings will enable SI to better capture the user's subtle intents.
However, evaluating these effects is difficult.
SI systems are usually evaluated by a human-centred qualitative approach, by observing the utility and effectiveness of the application for end-users. 
This approach has drawbacks in terms of replicability, scalability, and objectiveness, which makes it hard to perform convincing contrast experiments between different SI models. 
To tackle this problem, we explore a quantitative algorithm-centered analysis as a complementary evaluation approach, by simulating users' interactions and calculating the accuracy of the learned model. 
We use these methods to compare word-embeddings to bag-of-words features for SI.
}


\begin{document}
\firstsection{Introduction}
\maketitle
\label{sec:intro}
\pagenumbering{gobble}

Semantic interaction (SI)~\cite{Endert:he, Endert2015} is an interaction technique for non-experts to interact with the underlying algorithms of visual analytics (\textit{VA}) systems~\cite{cook2005illuminating}.
With SI, analysts can focus on reasoning and manipulating data in the 2D spatialization instead of directly interacting with the underlying machine learning (\textit{ML}) models~\cite{amershi2012regroup}.
It is, therefore, the system's responsibility to tune the ML models by capturing users' interactions and inferring the analyst's intent~\cite{self2016bridging}.
These intents are then translated to updated model parameters via semi-supervised machine learning techniques (ML)~\cite{Endert2015}, such as metric learning~\cite{6400486}. 
Hence, analysts can remain concentrated on their sensemaking activities~\cite{pirolli_2005}.

In this paper, we investigate two questions about SI:  
(1) Can word embeddings help SI learn user's interactive intents better than traditional features?
(2) How can we comparatively evaluate alternative SI models, such as in question (1)?

The human-centred approach~\cite{boukhelifa2018evaluation} is the primary evaluation method for SI systems.
The SI system is measured by the utility and effectiveness of the application for end-users through studies of human subjects. 
For example, ForceSPIRE~\cite{Endert:2012:SIV:2207676.2207741} is a visual text analysis system powered by SI, and its effectiveness was measured by analysts' performance on an intelligence analysis task.
This approach is highly dependent on human feedback, which leads to several challenges when comparing the effectiveness of different SI models because: 
 user interactions are difficult to precisely replicate;
 SI interactions are incremental and iteratively build;
 does not scale well to compare many model alternatives as needed in ML;
 and usage differences can mask subtle model performance differences.

To cope with these problems, we design and perform an additional quantitative algorithm-centred analysis that simulates human interactions as a complement to human-centered evaluation. In the evaluation, we use the labelled text datasets as the ground truth, since there is no ground truth for the user’s complex intents and concepts. 
Then the ground truth is used to simulate user's interactions, and we calculate and compare the accuracy of trained models. 

\begin{figure*}[htb]
  \includegraphics[width=\textwidth]{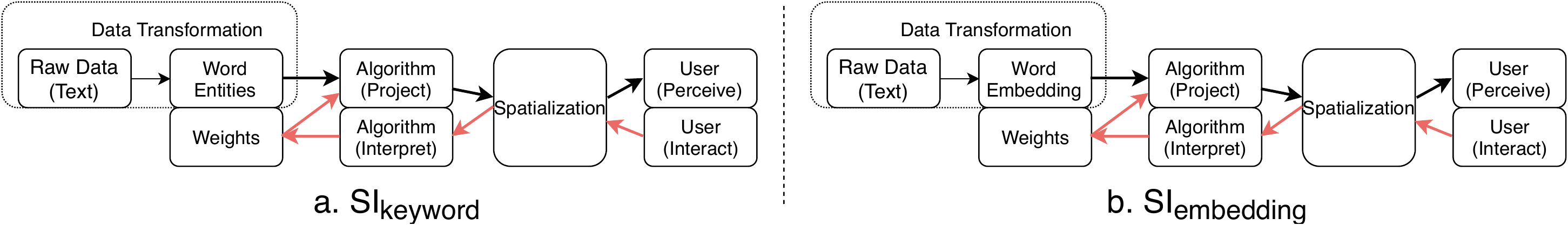}
  \caption{
  In the SI pipeline, distance metric learning interprets users' interactions on the projection. (a) In $SI_{keyword}$, the extracted features of text data are keywords; (b) In $SI_{embedding}$, the features are embedding vectors.}
  \label{fig:SI_pipeline}
\end{figure*}

In this paper, we first describe two alternative SI models for visual text analysis using different data features as inputs (embedding vectors, and keywords vectors). 
We then implement a generalized visual text analysis prototype with SI that can take both embedding vectors and keywords vectors as features. 
We then demonstrate how the two kinds of evaluation methods together provide a complementary evaluation and comparison of these two SI models.

\section{Semantic Interaction with Word Embeddings}
\label{sec:semantic-interaction}

We investigate the opportunity of using word embeddings to better support SI. SI's ability to infer the analyst's reasoning process is fundamentally limited by the feature space of the underlying machine learning. 
Text analytics systems with SI, such as ForceSPIRE~\cite{Endert:2012:SIV:2207676.2207741, 7042492} and Cosmos~\cite{DOWLING201949},  typically use keywords (such as text terms and phrases) as data features ($SI_{keyword}$ in Fig.~\ref{fig:SI_pipeline}a),  known as bag-of-words~\cite{zhang2010understanding}. 
Recently, word embedding techniques~\cite{mikolov2013distributed,collobert2008unified, 8416973, Peters:2018}, also known as deep learning representations~\cite{LeCun:2015dt, athiwaratkun2015feature, 7797053, Yosinski:wc}, have shown significant advantages in numerous tasks in natural language processing and information retrieval~\cite{Kusner:2015:WED:3045118.3045221,Socher13parsingwith,Le:2014:DRS:3044805.3045025} over bag-of-words features.
Therefore, the combination of SI with word embedding ($SI_{embedding}$ in Fig.~\ref{fig:SI_pipeline}) might enable better learning ability than $SI_{keyword}$ to update the underlying machine learning models~\cite{mikolov2013efficient}. The intuition is that word embeddings could represent more abstract concepts that are closer  to modeling human cognitive reasoning.
To test the hypothesis, we focus on the comparisons of $SI_{embedding}$ with $SI_{keyword}$ through the same SI pipeline but with different document features as input: 
\begin{itemize}
    \item \textbf{Keyword features used in $\mathbf{SI_{keyword}}$:} 
    We use TF-IDF values as the keyword features,  
    and word hashing to compress the large number of words from the document collection into 300 dimensions. 
        
    \item \textbf{Embedding features used in $\mathbf{SI_{embedding}}$}:
    We use the pre-trained GloVe model~\cite{pennington2014glove} with 300 dimensions to extract embedding features,  
    and use the ``basic averaged word embeddings'' method to compute average word embedding for all words in a document~\cite{Le:2014:DRS:3044805.3045025}. 
\end{itemize}

\subsection{Application Prototyping}
\label{sec:implementation:prototyping}

As shown in Fig.~\ref{fig:SI_pipeline}, $SI_{embedding}$ has a similar structure with $SI_{keyword}$ as they both use numerical vectors to represent documents (embedding vectors, and TF-IDF vectors). 
We are able to switch between $SI_{keyword}$ and $SI_{embedding}$ in the same SI prototype. 
We build the prototype upon the foundation of Andromeda~\cite{self2018observation}, a visual analytics tool for exploring high-dimensional data projections.
The prototype can use either bag-of-words or embedding vectors as features and update the feature weights to capture analysts' intents. 
For bag-of-words, the system will up-weight the shared words in the dragged documents in response to the analyst dragging two or more documents closer together. 
For embedding vectors, the system will up-weight the dimensions of the embedding vectors with similar patterns in the dragged documents.

\begin{figure*}[h!]
  \includegraphics[width=\textwidth]{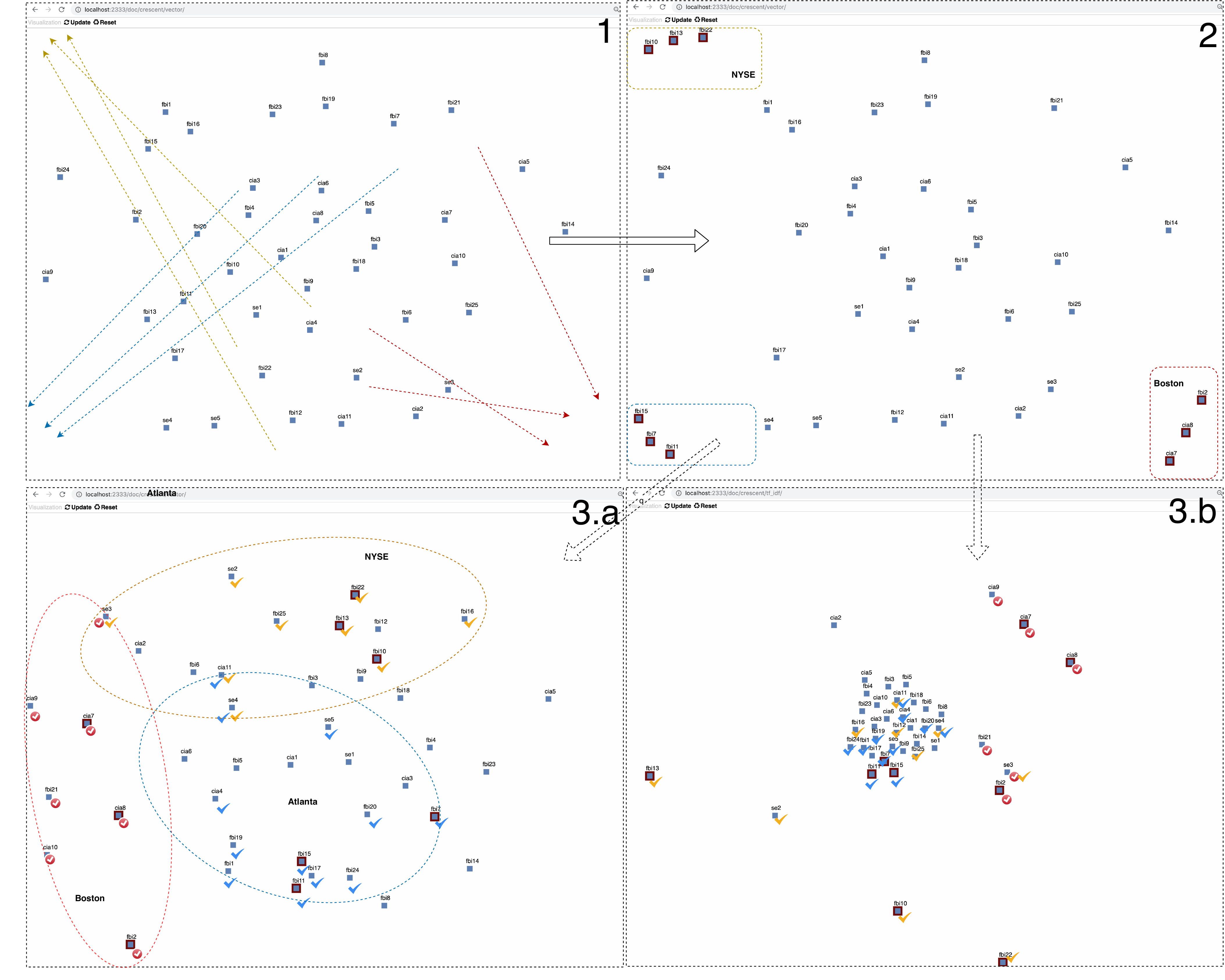}
  \caption{Several screenshots during the two case studies, as discussed in Section~\ref{sec:case-study}: 
  \textbf{Frame 1 and 2} show the similar initial steps performed by analysts in both case studies. 
  \textbf{Frame 3.a} shows the resulting projection based on analysts' interactions, in the case study using $SI_{embedding}$. 
  \textbf{Frame 3.b} shows the resulting projection based on analysts' interactions in the case study using $SI_{keyword}$.
  }
  \label{fig:case-study}
\end{figure*}

As shown in Figure~\ref{fig:case-study}-1, the visualization is a projection of documents. 
The distance between documents reflects their relative similarity according to a weighted distance metric over their features.
At first, documents scatter in the workspace based on their features and an equally-weighted dimension reduction of the high-dimensional representation. 
If two documents are positioned close to each other in this initial projection, it implies that these two documents are similar based on all vector features. 
Semantic interaction enables analysts to directly manipulate the projection by 
moving the documents to express their own
domain knowledge about desired similarities.
With this interaction, analysts can express the semantic relationships between documents, which thereby informs the underlying weighted distance model and updates the projection. Through this, the projection can be customized according to the learned intent of the analyst. 
 
\begin{table*}[tb]
  \centering
  \begin{tabular}{|l||c|c|c|c|c|c|c|}
    \hline
    \textbf{Model} & \textbf{$T_{rec}$} & \textbf{$T_{religion}$} & \textbf{$T_{sys}$} & \textbf{$T_{vis}$} \\
    \hline
    \hline
    $SI_{embedding}$ & (\textbf{0.921}, \textbf{0.812}) & (\textbf{0.829}, \textbf{0.773}) & (\textbf{0.895},  \textbf{0.774}) & (0.958, \textbf{0.809})\\
    \hline
    $SI_{keyword}$ & (0.497, 0.570) & (0.576, 0.581) & (0.511, 0.584) & (\textbf{0.961}, 0.793)\\
    \hline
  \end{tabular}
  \caption{Accuracies of $SI_{embedding}$ and $SI_{keyword}$ on each of the four tasks~($T_{rec}$, $T_{religion}$, $T_{sys}$, and $T_{vis}$). Two kinds of accuracies are measured: the average accuracy of the trained model after the last interaction and the average accuracy of the models over all interactions.}
  \label{tab:result}
\end{table*}

\begin{figure*}
    \begin{subfigure}[tb]{0.245\textwidth}
        \includegraphics[width=\textwidth]{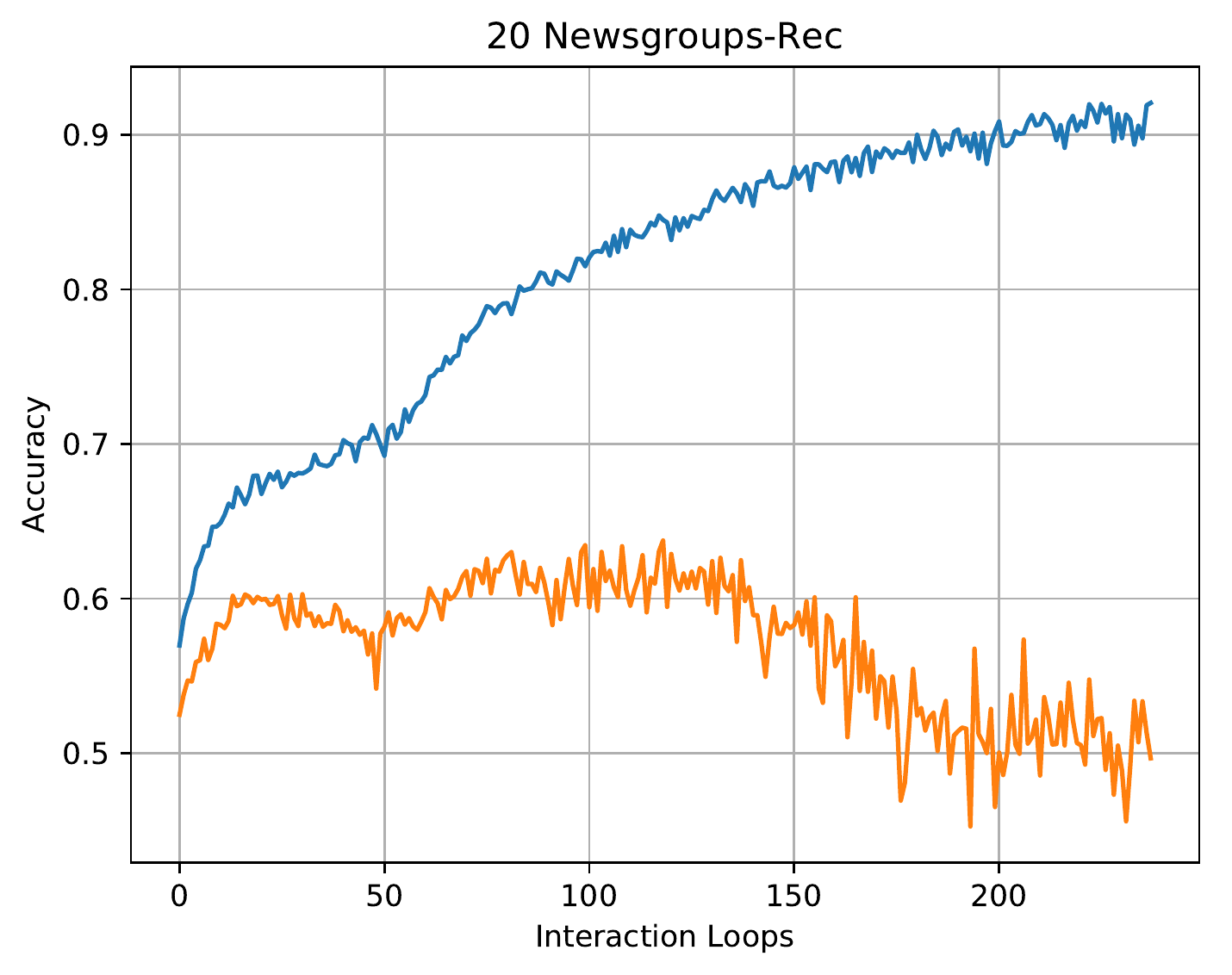}
        \caption{$T_{rec}$}
        \label{fig:rec}
    \end{subfigure}
     \begin{subfigure}[tb]{0.245\textwidth}
        \includegraphics[width=\textwidth]{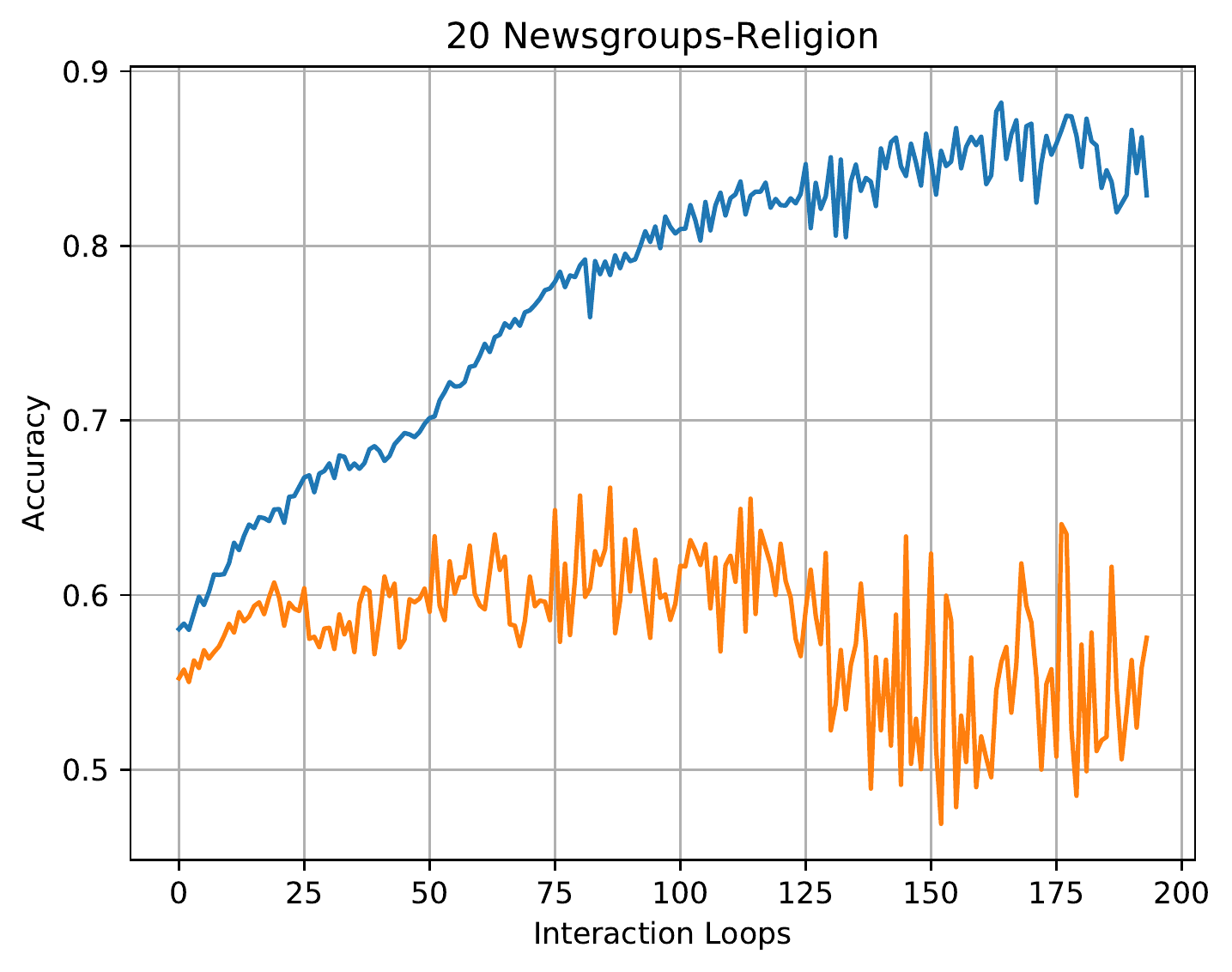}
        \caption{$T_{religion}$}
        \label{fig:religion}
    \end{subfigure}
     \begin{subfigure}[tb]{0.245\textwidth}
        \includegraphics[width=\textwidth]{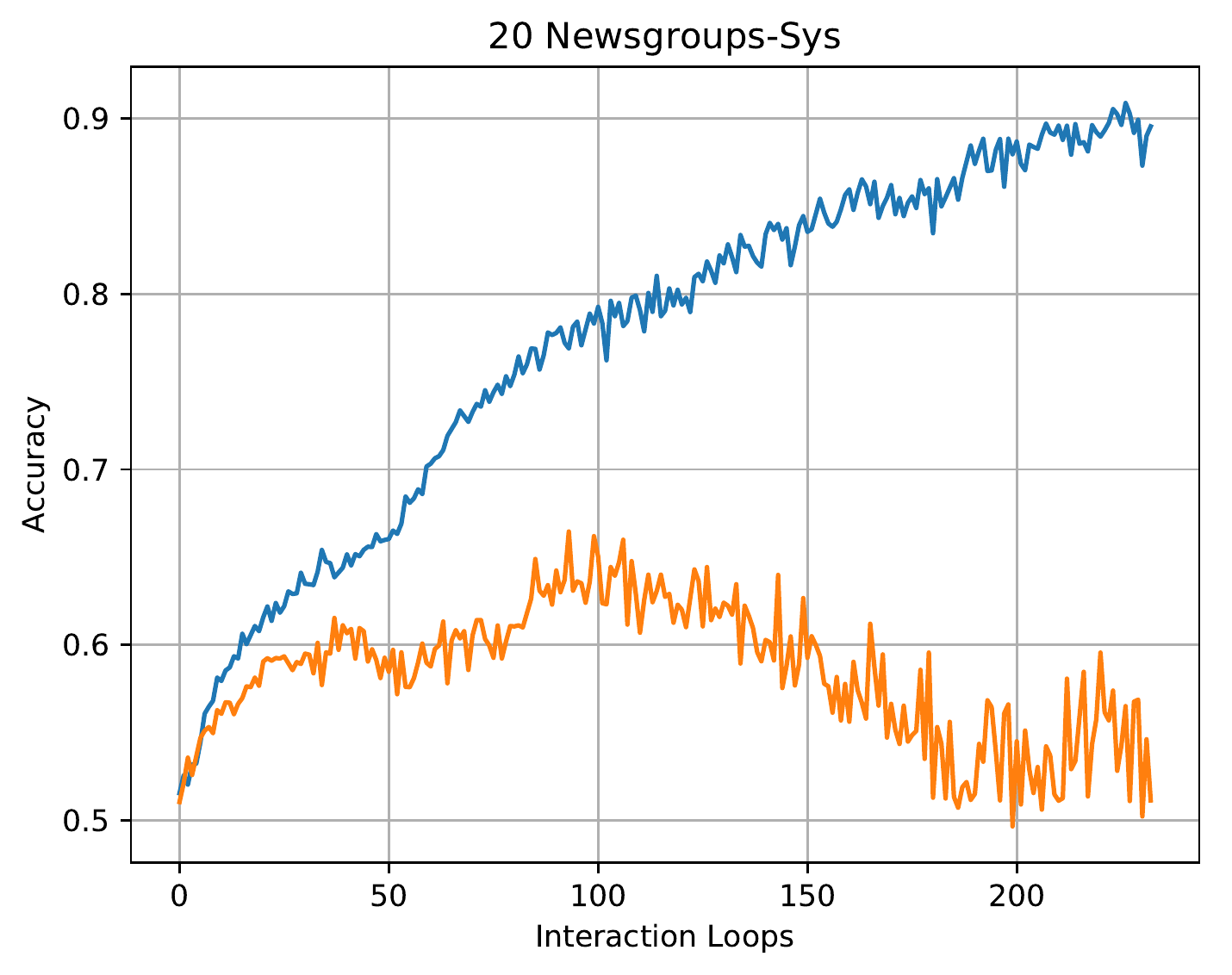}
        \caption{$T_{sys}$}
        \label{fig:sys}
    \end{subfigure}
     \begin{subfigure}[tb]{0.245\textwidth}
        \includegraphics[width=\textwidth]{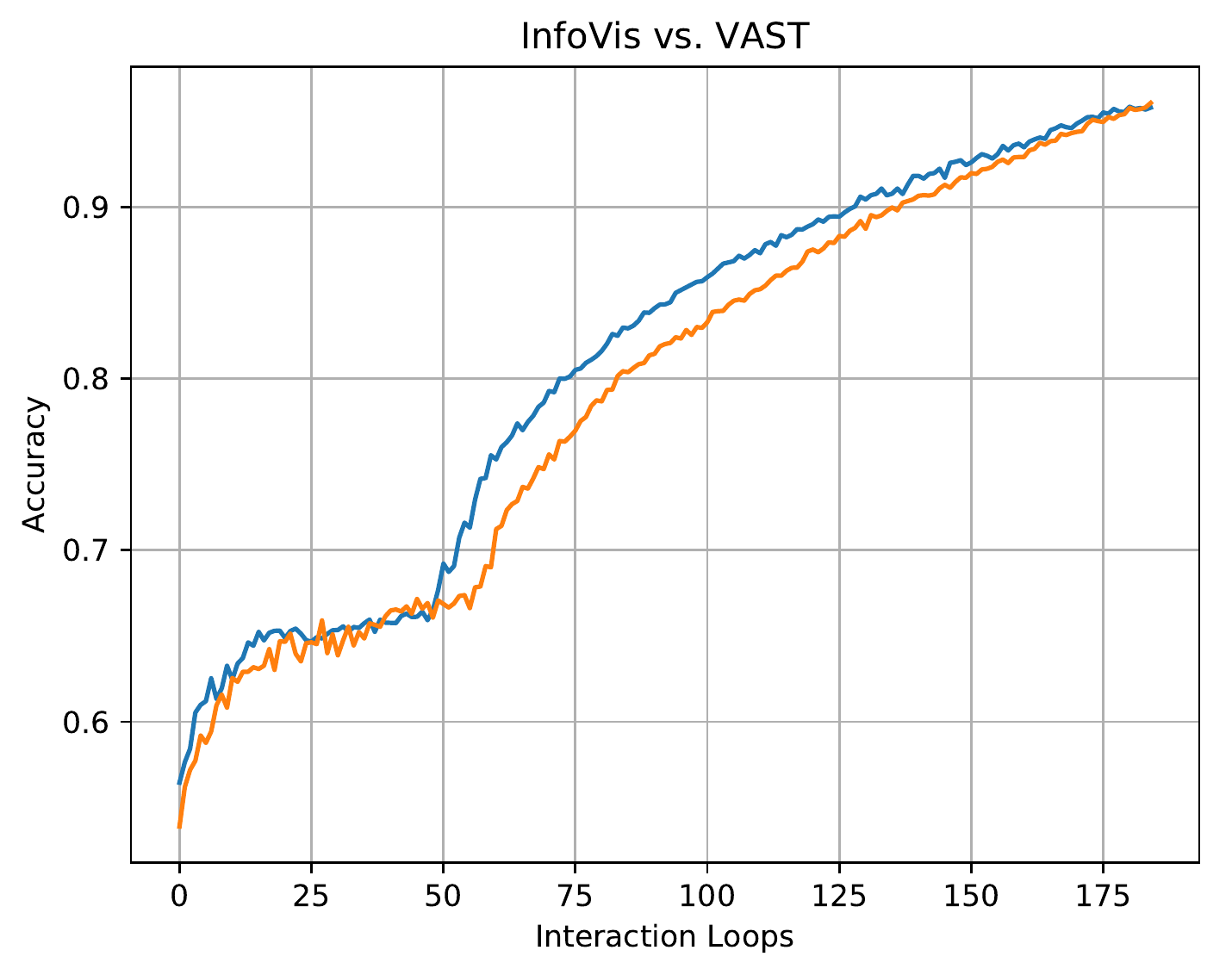}
        \caption{$T_{vis}$}
        \label{fig:ieeevis}
    \end{subfigure}
    \caption{
    The accuracies of both $SI_{embedding}$ (blue) and $SI_{keyword}$ (orange) over each interaction across the four tasks (\small{$T_{rec}$, $T_{religion}$, $T_{sys}$, and $T_{vis}$).}}
    \label{fig:result}
\end{figure*}

\section{Evaluating SI Embedding}
Traditionally, the human-centred approach is a good method to test if the SI model is an effective application for analysts to perform sensemaking tasks. 
However, comparing two SI models ($SI_{keyword}$ and $SI_{embedding}$) through this approach is inadequate.
Thus, we also design a replicable quantitative study to validate $SI_{embedding}$ in comparison to $SI_{keyword}$ in 4 simulated text analysis tasks of different difficulty levels.

\subsection{User-Centered Qualitative Analysis}
\label{sec:case-study}

In this section, we present a practical use case about intelligence analysis as our qualitative study to compare $SI_{keyword}$ against $SI_{embedding}$.
For this analysis, we engaged an expert in intelligence analysis to review the visual result generated from both $SI_{embedding}$ and $SI_{keyword}$ to provide qualitative feedback, and provide grounding for the quantitative analysis.

\subsubsection{Crescent Dataset}
The crescent dataset~\cite{crescent} has 42~fictional intelligence reports regarding a coordinated terrorist plot in Boston, New York, and Atlanta.
Only 24~reports are relevant to these plots.
The task for our qualitative analysis is to identify these three terrorist threats by using $SI_{embedding}$ or $SI_{keyword}$. 

By moving documents in the visualization according to their perceived similarity, analysts express their reasoning process to the system.
The ground truth of this task is as follows, which we use to measure the accuracy of the models: 
    \textbf{Boston:} cia7, cia8, cia9, cia10, cia11, fbi1, fbi2, fbi21, se3, se4; 
    \textbf{New York:} cia11, fbi1, fbi10, fbi13, fbi16, fbi22, fbi25, se2, se3, se4;
    \textbf{Atlanta:} cia4, cia11, fbi1, fbi7, fbi11, fbi15, fbi17, fbi19, fbi20, fbi24, se4, se5;
    \textbf{Irrelevant:} fbi3, fbi4, fbi5, fbi6, fbi8, fbi9, fbi14, fbi18, fbi23, cia1, cia2, cia3, cia5, cia6, se1.

\subsubsection{Case study with $SI_{embedding}$}
In this first case study, the analyst performs the task by using the prototype system with embedding as the document features. 
As shown in Figure~\ref{fig:case-study}-1, the analyst updates the layout to reflect the perceived similarities between the documents, grouping three documents about ``New York'' to the top left region of the projection, indicated by the yellow arrows in Figure~\ref{fig:case-study}-2, three documents about ``Atlanta'' to the bottom left region indicated by blue arrows, and three documents about ``Boston'' to the bottom right part indicated by red arrows. 

After the layout updates, some semantic relationships are revealed (Figure~\ref{fig:case-study}-3.a). 
The left cluster in red circle contains documents about ``Boston,'' the top right cluster in yellow circle contains documents about ``NYSE,'' and the bottom right cluster in blue circles contains documents about ``Atlanta.'' 
We found that there are semantic mappings between this updated layout and the ground truth;
all relevant reports that belong to single plot are well placed into each cluster. 
Reports about the coordination between two or more plots are between clusters.
For example, the report ``se3'' contains information regarding both the ``Boston'' and ``NYSE'' plots, so the document is located between these two clusters. 
Thus, $SI_{embedding}$ can capture analysts’ intents and update the model features accordingly. The updated layout then reflects the semantic meanings behind the intents.

\subsubsection{Case Study Using $SI_{keyword}$}
Mirroring our case study with $SI_{embedding}$, the analyst performs similar interactions with the prototype.
As shown in Figure~\ref{fig:case-study}-3.b, from the updated layout, there are no clear boundaries between different plots. 
Even after continued interactions, the model is unable to properly differentiate between the three terrorist plots. 
For example, documents about ``Atlanta'' and ``NYSE'' are mixed together.
Furthermore, the  documents that are pulled out of the big central cluster in panel 4 are primarily the user interacted documents from panel 2. This might indicate that the model is over-fitting based on some specific unimportant keywords.

\subsubsection{Expert Review}
Besides comparing with the ground truth as we mentioned in Section~\ref{sec:case-study}, we also asked an expert familiar with the dataset to evaluate the updated layout generated in both $SI_{embedding}$ and $SI_{keyword}$ (as shown in Figure~\ref{fig:case-study}-3.a and Figure~\ref{fig:case-study}-3.b).
The expert noted that the $SI_{embedding}$ layout provides more meaningful information about there plots than $SI_{keyword}$ layout because the documents from different plots are grouped in different regions.
In contrast, the layout generated by $SI_{keyword}$ makes it difficult to clearly distinguish between these plots. 
For example, documents in the the $SI_{embedding}$ layout regarding coordination between the plots are placed between the relevant groups.
This is exemplified by how the report ``se3'' is between the ``NYSE'' group and the ``Boston'' group since it discusses how terrorists involved in both plots communicated.
In the layout produced by $SI_{keyword}$, this document lies at the bottom right of the projection, which does not have such an immediately apparent and relevant semantic meaning.

Additionally, several irrelevant reports were distinguished from the groups in the layout produced by $SI_{embedding}$, such as ``cia5'', ``fbi23'' and ``fbi14,'' even though they share many keywords with other relevant reports. 
In contrast, the layout produced by $SI_{keyword}$ only causes one irrelevant report (``cia2'') to be pulled away from other reports. 

\subsection{Algorithm-centered Quantitative Analysis}
\label{sec:quantitative-analysis}

In this subsection, we describe our quantitative comparison between $S_{keyword}$ and $SI_{embedding}$.
We begin with the experimental design, followed by a description of
the datasets we choose for the experiment. 
Finally, we show the evaluation results and discuss the SI inference abilities of the two models.

\subsubsection{Experiment Setup}
We need ground truth to evaluate the underlying algorithms used in $SI_{keyword}$ and $SI_{embedding}$ quantitatively.
There is no ground truth about users' complex intents and interactions, such as a labeled intelligence hypothesis in a 2D projection.
However, there are sufficient text datasets labelled for classification, such as 20newsgroup (section~\ref{sec:evaluation:quantitative-analysis:dataset}).
The labels can be used to simulate different positions in the projection: for example, documents with negative label can be located at the top left part of the spatial layout, and documents with positive label can be located at the bottom right part. 
Therefore, we can use classification datasets mapped in the 2D projection as the ground truth, and we only evaluate the inference ability to capture the intents embedded in classification tasks. 
Then, subsets of the  documents from different positions in the 2D projection will be picked and used as simulated semantic interactions from analysts.

Furthermore, since \textit{incremental formalism}~\cite{Shipman1999} is an important aspect of semantic interaction, we must test the models' ability to learn incrementally over the course of many interactions.
To simulate and evaluate the \textit{incremental inference} ability  of these models, the selected interactions will be iteratively passed into the SI model.
Through these iterative SIs, the underlying models incrementally update to display better results to the analyst. 

For example, in the task $T_{vis}$ (sec.~\ref{sec:evaluation:quantitative-analysis:dataset}), we simulate that the analyst wants to organize documents by separating the two conferences (InfoVis and VAST) in the 2D projection as ground truth: InfoVis documents placed at top left of the projection, and VAST documents placed at bottom right. 
Then in each loop, five documents from InfoVis collection, and five documents from VAST collection will be picked and simulated as semantic interactions: moving five documents from InfoVis to $(0, 0)$ position and moving another five documents from VAST to $(1, 1)$ position. 
In each iteration, the SI model ($SI_{embedding}$ or $SI_{keyword}$) is executed to infer updated  model parameters from the simulated semantic interactions. 
Then the inference ability of each model can be calculated by the quality of the underlying model in classifying InfoVis documents from VAST based on the ground truth.

\subsubsection{Datasets and Tasks}
\label{sec:evaluation:quantitative-analysis:dataset}
To perform this quantitative analysis, we use two datasets: 20News dataset~\cite{Lang95} and Vispubdata dataset~\cite{Isenberg:2017:VMC}.
The 20 Newsgroup dataset is a collection of newsgroups posts on 20 topics. 
Based on this dataset, we create three tasks to simulate users' intents.
Each of the 3 tasks are to separate the documents into 2 groups based on 2 pre-determined topics:
    \textbf{$T_{rec}$}: 594 documents from ``rec.autos'' and 600 documents ``rec.motorcycles''; 
    \textbf{$T_{sys}$}: 578 documents from ``comp.sys.mac.hardware'' and 592 documents from "comp.sys.ibm.pc.hardware";
    \textbf{$T_{religion}$}: 379 documents from ``talk.religion.misc'' and 599 documents from ``soc.religion.christian''.

The Vispubdata dataset contains information on IEEE Visualization (IEEE VIS) publications. 
We select the abstracts of the academic papers published in two conferences from the dataset: InfoVis (IEEE Information Visualization), and VAST (IEEE Visual Analytics Science and Technology).
The documents from different conferences are used to represent different concepts (users' intents),
We create one task based on these abstracts to separate InfoVis papers from VAST papers, defined as $T_{vis}$, including 397 papers from InfoVis and 531 papers from VAST.

\subsubsection{Results}
In this experiment, we have run the $SI_{embedding}$ and $SI_{keywrod}$ model multiple times on the four tasks with ground truth. 
After each interaction (involving  five document movements), we calculate the model accuracy using the k-nearest-neighbour (kNN) classifier~\cite{Cover:2006:NNP:2263261.2267456} (as done in Dis-Function~\cite{6400486}): using the cross-validation over the data and set k to 3.
The average accuracy of the two models over all iterations for fours tasks is shown in Figure~\ref{fig:result}.

Table~\ref{tab:result} shows the average accuracy of the last interaction (performed multiple times) and average accuracy of $SI_{embedding}$ and $SI_{keyword}$ across the four tasks.
$SI_{embedding}$ achieved the highest scores in the first three tasks, indicating that $SI_{embedding}$ provides more accurate representations of the documents after the interactions, indicating its higher inference ability over $SI_{keyword}$. 
In $T_{vis}$, $SI_{embedding}$ and $SI_{keyword}$ have similar accurracies, meaning their inference abilities were roughly equal in this task. 

As shown in the Figure~\ref{fig:result}, we further analyzed the incremental updates from the two models by evaluating each model's accuracy after every iteration, thereby simulating the incremental analysis process. 
In the first three tasks, it is shown that using $SI_{embedding}$ can get better performance than $SI_{keyword}$ over all the iteration loops. 
The accuracy of $SI_{embedding}$ generally increases with each interaction, showing  incremental changes to the model's representation of the user's intent over time.
In contrast to the first three tasks, however, in $T_{vis}$, both $SI_{emebedding}$ and $SI_{keyword}$ have similar accuracy over the interactions.
Finally, the desired accuracy should be as close to 1 as possible. 
$T_{vis}$ is a relatively ``easy'' task, as reflected by the final accuracies of both models close to 1.0.
These results indicate that $SI_{keyword}$ performs well in relatively easy tasks. 


\subsection{Discussion}
Our experimental results substantiate our hypothesis that $SI_{embedding}$ is more effective than $SI_{keyword}$ for modeling user intent, and better supports incremental formalism based on users iterative interactions. The human-centered qualitatively evaluation method shows $SI_{embedding}$ is more effective than $SI_{keyword}$ in real world analysis tasks. 
This provides a direct evaluation of SI models by the observed utility and effectiveness for end-users. 
The quantitative analysis offers  a complementary approach, and provides replicable and scalable evaluations for SI models. 
It provides more stable and detailed feedback about the SI model performance in tasks of different difficulty levels. 
Thus, we confirmed that word embedding can better support SI by better capturing the users' high-level interactive intents.

\section{Conclusion}

In this work, we presented $SI_{embedding}$ as an alternative  to the traditional bag-of-words model ($SI_{keyword}$) often used in visual text analytics systems. 
To make a complete and convincing comparison between $SI_{embedding}$ and $SI_{keyword}$, we performed a quantitative evaluation by simulating analysts' interactions and  calculating the accuracy of the underlying trained ML models, as a complement to the traditional user-centered qualitative evaluation. Results indicate that deep learning distributed representations, such as word embedding, can be exploited to improve interactive visual analytics methods such as semantic interaction.


\bibliographystyle{abbrv-doi}

\bibliography{template}
\end{document}